\documentstyle[epsf]{l-school}  
   
\begin{document}   
\markboth{K. D. Kokkotas}{Pulsating Relativistic Stars}   
\setcounter{part}{1 }   
\title{Pulsating Relativistic Stars}   
   
\author{K. D. Kokkotas}   
\institute{Department of Physics,\\   
           Aristotle University of Thessaloniki,\\   
           Thessaloniki 54006, Greece
\footnote{
Review talk for the proceedings of the Les Houches School on
{\em Astrophysical Sources of Gravitational Waves}        
Eds J-A.~Marck and J.~P.~Lasota, Springer-Verlag (1996)}           
           }   
\maketitle   
   
\section{INTRODUCTION}   
   
Pulsating stars have given important information in classical astronomy and 
are   still an active research field, {\em e.g.} {\em asteroseismology} or 
{\em helioseismology}.  
By studying the rich oscillation stellar spectra, optical astronomy    
makes predictions for the stellar equations of state (EOS), 
the evolutionary age of the stars, and in the case of
Cepheids one can make exact predictions for the distance of the object.   
   
For relativistic astrophysics the pulsations of relativistic stars are 
of great importance since through them one can discuss stability 
properties of compact objects and, moreover, pulsating stars can be a 
promising source of gravitational waves.  
It is expected that during the first few seconds which will   
follow a supernova explosion 
the newly formed neutron star will pulsate wildly and this pulsation   
will be mainly damped due to the emission of gravitational radiation which   
will carry away the characteristic signature of the collapsed object. 
The energy stored in the pulsation will be of the same order as the 
kinetic energy of the collapse and in this way a significant part of the 
original mass-energy of the newly formed neutron star will be radiated away 
as gravitational radiation. 
With the present sensitivity of the resonant detectors it is quite sure that 
the gravitational waves will be detected if a supernovae explosion takes 
place in the Local Group of galaxies. 
Depending on the details of the collapse \cite{BM94}, with the future 
generation of resonant detectors it is possible that events up to the 
distance of the Virgo cluster to be detectable.

The study of stellar pulsations in General Relativity (GR) has been 
initiated nearly thirty years ago by  Thorne and his collaborators 
\cite{TC67,Thorne68,PT69,Thorne69a,Thorne69b}, and was intended  
mainly in the calculation of the $f$ (fundamental) 
mode since this is the mode through which    
most of the gravitational radiation of the star is radiated away.   
All the other {\em fluid modes}, $g$ (gravity), $p$ (pressure),   
$s$ (shear), $t$ (toroidal), $i$ (interface) modes, can be calculated with    
quite high accuracy with the Newtonian dynamics since they don't emit  
significant amounts of gravitational radiation. 
This Newtonian like picture has changed significantly in the last    
ten years due to some new ideas by Kokkotas and Schutz 
\cite{KS86, KS92} and   
Chandarsekhar and Ferrari \cite{CF91a} but also due to subsequent work by  
Andersson and Kojima.  
The new approach has been based on the  fact that the stellar   
pulsations in GR should not be treated in a semi-Newtonian manner   
because then some important features of the theory will be overlooked   
since they are related   
uniquely to the relativistic nature of the problem and they do not   
exist in the Newtonian theory. These unique new features are:  
\begin{itemize}   
\item     the existence of a new set of modes, the gravitational   
          $w$ave modes ($w$-modes),    
          which are purely spacetime modes and which cannot   
          be seen in Newtonian theory    
          \cite{KS86, KS92, Koj88, AKS95, KAK95, AKK96, AKS96, LNS93}.   
\item     that the stellar pulsations can be treated as a scattering 
          problem \cite{CF91a} {\em i.e.} in the perturbation equations 
          only the variations of the gravitational field   
          will enter while  the fluid perturbations will be absent. 
          This approach   
          exists even in the limit of Newtonian stars \cite{FG94,CF95}.   
\item     that in the slow rotation limit the polar modes excite the  
          axial ones due to the dragging of the inertial frames  
          \cite{CF91c,Koj92,Koj93,Koj93a}  
\item     that the axial  modes which were 
          thought  not to exist for non-rotating stars    
          do exist \cite{CF91b} and that they show a spectrum similar   
          to the $w$-modes \cite{KDK94} and that, more or less,   
          they can be considered on an equal footing as the polar   
          modes \cite{KAK95,AKK96}.   
          Recently, it has been shown \cite{AK96} that these modes can  
          be excited and emit as much energy as an oscillating black-hole.   
\end{itemize}   
   
These new developments in the last decade in the theory of relativistic   
stellar pulsations will be the subject of the following sections.   
We will not discuss stability since for non-rotating stars there is 
no significant progress in the last ten years and thus the reader can 
refer to a review article by Schutz \cite{Schutz87}.

\section{FLUID MODES}   
   
The fluid modes are modes that exist in both Newtonian and GR  
treatment of the stellar pulsation and their main characteristics do not   
depend on the specific gravitational theory.   
If we consider an  unperturbed fluid it will be described in both theories  
by the mass and momentum conservation equations and an equation which  
describes the gravitational field (Poisson or Einstein equations).  
By assuming a  small variation from sphericity we can 
practically describe all   
fluid and dynamical quantities in terms of spherical harmonics while a   
harmonic dependence on time can also be used.   
In Newtonian theory all the quantities which describe 
the variations of both the   
fluid and the potential are scalar quantities 
and their decomposition in spherical harmonics has only one parity.   
In GR, the variations of the spacetime metric are 
tensor quantities and thus their decomposition in spherical harmonics 
has two parities, {\em odd} and {\em even} ones.   
The {\em even} parity or {\em polar} perturbations are similar 
to the Newtonian   
ones and describe {\em spherical} deformations while the {\em odd} parity   
or {\em axial} ones (called also {\em toroidal})  
are degenerate in Newtonian theory. 
They can set the star into a continuous non-varying differential rotation.   
Nevertheless, this degeneracy is erased in the presence of rotation,   
magnetic fields, or non-zero shear modulus.   
Thus, in principle, in the absence of rotation in both  
theories these modes are   
unimportant because they cannot describe time dependent motions and as   
a result there is no emission of gravitational radiation.   
We shall discuss these modes in the next section where we shall 
see that the  
Newtonian approach to the problem is not complete.   
The perturbation equations have the general form   
\begin{equation}   
{\cal D}_1 W + F\left( W',Z',W,Z;\ell,\omega \right) = 0 
\label{pert1}   
\end{equation}   
\begin{equation}   
{\cal D}_2 Z + G\left( W',Z',W,Z;\ell,\omega \right)= 0 
\label{pert2}   
\end{equation}   
where $W(r)$ is a function of  
the radial, $\xi_r$, and tangential, $\xi_\theta$, components of the 
fluid displacement vector.  
$Z(r)$ is a function of the components which describe the  
variation of the gravitational field  
\footnote{ 
In the Newtonian theory they are the variation of the potential 
$\delta U$ and 
its derivative, while in GR they are components of the perturbed part of 
the metric, {\em e.g.}  for a certain choice of the gauge,  $h_{tt}$ and 
$h_{\theta \theta}$. 
In both cases the system of differential equations is of 4th order 
\cite{DL85}, in the gauge used in \cite{CF91a} the system is of 5th order 
but it has been proven \cite{IP91} that by a gauge transformation it can always 
reduced to a 4th order one. A gauge invariant presentation by Moncrief
\cite{Moncr74} has never been used up to now.
}. 
$\ell$ is the harmonic index   
($m$ is not present because we have a $(2\ell+1)$-fold degeneracy due to   
the rotational symmetry of the equilibrium structure around any arbitrary   
axis) and $\omega$ is the pulsational frequency.   
${\cal D}_1$ and ${\cal D}_2$ are  
both  wave operators in GR with propagation velocities, the acoustic one
for the first and the speed of light for the second equation
(in Newtonian theory ${\cal D}_2$ 
is not a wave operator thus we have no emission of gravitational waves). 
In GR outside the star the equation (\ref{pert2}) reduces to either 
Zerilli (polar) or Regge-Wheeler (axial) wave equation 
\cite{Chandra83}  depending on the parity.  
The equations (\ref{pert1}) and (\ref{pert2}) together with a set  
of appropriate boundary conditions formulate an eigenvalue problem  
with $\omega$ being the eigenvalue and $W(r)$ and $Z(r)$ the corresponding  
eigenfunctions.  
The boundary condition for the existence of quasi-normal modes 
(QNM) is that  
at infinity there is no incoming gravitation radiation.  
In this way a discrete spectrum of frequencies for a certain 
stellar model can  be revealed.  
   
The richness of the spectrum depends on the complexity of the   
stellar model and each new feature that we add to the stellar 
model results in a new family of modes. 
\footnote{ 
The naming of the various families of fluid modes comes from 
a monumental work  by Cowling in early '40s \cite{Cowling41}. 
} 
The simplest possible model for a neutron star is a non-rotating fluid ball   
at zero temperature. 
If the fluid has constant density this   
model   supports  only one pulsation mode for each multipole $\ell$.   
This  mode, which in Newtonian theory has oscillation frequency   
\begin{equation}   
 \omega (R^3/M)^{1/2}= \sqrt{ 2\ell (\ell-1)/(2\ell+1) } \approx 0.894 
 \quad {\rm for}\ \ell=2 \ ,   
\label{fmode}   
\end{equation}    
where $R$ and $M$ are the radius and mass of the star in units where   
$c=G=1$, was first studied by Kelvin \cite{Kelvin} and is usually referred   
to as the $f$-mode  (fundamental mode).   
It is distinguishable by the fact that there are no nodes in the   
corresponding eigenfunctions inside the star.   
In a way, the $f$-mode is due to the interface between the star   
and its surroundings.   
The eigenfunctions of such modes would typically have   
maxima at the interface and fall off away from it.   
This is exactly the character of the $f$-mode: It   
reaches maximum amplitude at the surface of the star (see for example   
Figures 7--8 in \cite{KS92}).   
A comprehensive set of quadrupole $f$-mode calculations has 
been carried out   
for 13 equations of state by Lindblom and Detweiler \cite{LD83}.   
The frequencies are in the range of 1.5-3.5 kHz and the damping 
times are of   
the order of 0.1-0.5 secs, {\em i.e.} the star undergoes a few hundred 
oscillations  before it damps out.  
   
A somewhat more realistic star consists of a perfect fluid. Then one   
must specify the equation of state, and most studies to date have been   
for simple polytropes.   
For this stellar model a second set of modes --  the $p$-modes,   
the restoring force of which is pressure -- exists.   
The  oscillation frequencies of the $p$-modes depend   
directly on the travel time for an acoustic wave across the star.   
This would typically lead to higher oscillation frequencies starting   
from 5-6 kHz while their damping times are significantly longer than  
those of the $f$-modes and increase with the order of the mode.  
   
When the temperature of the star is non-zero a further set of modes   
comes into play.   
These modes are mainly restored by gravity and are   
consequently referred to as $g$-modes.   
For a star in convection, {\em i.e.,} when the entropy is constant,   
the $g$-modes are all degenerate at zero frequency.   
In general, however, their oscillation   
frequencies depend directly on the central temperature of 
the star and are typically smaller than a few hundred Hz.   
   
The three families of {\em fluid} modes discussed so far, 
the $f$-, $p$- and   
$g$-modes, all belong to the class of polar modes.   
For these models there are no {\em fluid} axial modes, since the   
stellar models discussed are all somewhat unrealistic.   
If even more realistic stellar models are considered, for example   
it is expected that neutron stars will have a kilometer-thick crust   
which will  crystallize, then, if   
the shear modulus in the crust is non-zero, axial modes do exist  
\cite{ST83,finn90}. 
There will be also families of modes directly   
associated with the various interfaces inside the star which will not 
be discussed here and for a detailed discussion the reader should refer 
to an excellent presentation in \cite{McD88}. 
   
The maximum pulsation energy is stored in the $f$-mode   
on which the fluid parameters undergo the largest changes, 
typical values are   
of the order of $10^{53}-10^{54}$ ergs \cite{LD83}.
If the gravitational radiation is the only damping mechanism of the 
pulsations (which is approximately true \cite{McD88}) then for the Newtonian
ones the emission 
rate can be calculated by the quadrupole formula \cite{Thorne69b, OH73} 
and the damping rate 
will be just $E/{\dot E}$ (for the analytic formula see \cite{BS82}). 
In GR approach the pulsation frequency is complex and the imaginary part 
corresponds to the damping of the pulsation. 
The calculations of the $f$-mode frequencies for the same mean density stars 
have shown that both Newtonian and GR predict more or less the same values.
This picture is also true    
for all the other families of fluid modes and thus they can be studied 
in Newtonian theory \cite{McD88}.   
The previous picture is partially true for the damping rate since when 
the star is fully  
relativistic ({\em e.g.} $M\approx 1.4M_\odot$ and $R\approx 10$Km) the  
Newtonian theory predicts 2-3 times faster damping of the oscillations   
\cite{BDLS85}. 
The post-Newtonian (pN) treatment of the problem reduces this gap 
significantly \cite{cutler91,CL92,KS96}  and it is a promising approach 
especially for the study of rotating stars. 
  
As it has been seen by equation (\ref{fmode}) the $f$-mode  
frequency scales with the mean density of the star, while its damping   
time (due to radiation reaction) depends on the compactness of the star,  
{\em i.e.} 
the more compact the star is the faster the oscillations damp out 
(for a discussion of the damping   
times including viscosity see \cite{CLS90}).  
These two properties of the $f$-mode frequency are 
useful for the estimation of the masses and radii 
of pulsating neutron stars. 
  
We shall conclude this section by mentioning a  
very useful approximation for the study of the fluid modes, the so called   
{\em Cowling Approximation} in which the perturbations of the potential 
(or the metric) are set to  zero and thus the whole problem is described  
with just one wave equation for the fluid.   
In this approximation both frequencies   
and damping times can be revealed with an error usually less than $10\%$   
\cite{Robe68, McDer83,finn88}.  
An additional advantage  is that  
the simple wave equation can be studied 
even analytically in the WKB approximation and the properties of the 
various families of modes   
described above can be visualized in an elegant way \cite{Unno89}.

\section{SPACETIME OR W-MODES}   
   
As it has been shown in the previous section the treatments of stellar   
pulsations in Newtonian theory and in GR do not differ significantly.   
Actually, when the pN correction were considered even the 
gap in the estimation  
of the damping times has been considerably reduced. 
Thus the natural question is what is actually the role of the equation   
(\ref{pert2}) being a wave one in GR?  
Is it just for the direct calculation of the damping  
times during the solution of the eigenvalue problem and not  
afterwards from the ratio $E/{\dot E}$? 
The answer is clearly no.   
Spacetime has it own dynamics and does not play a passive 
r\^ole as just the   
medium in which gravitational waves propagate; moreover, 
it has its own unique spectrum.  
  
Let us start by assuming a simplified version of the problem.  
We shall consider the Inverse of the Cowling Approximation 
(ICA) \cite{AKS96}  
which has been discussed earlier, 
{\em i.e.} instead of {\em freezing} the spacetime  
perturbations we shall {\em freeze} the fluid ones. 
Then the whole problem is 
described by just one second order wave equation both inside 
and outside the  
star (as for the axial perturbations), 
which together with the appropriate boundary conditions (regularity of 
the perturbed quantities in the center and only outgoing waves at infinity)  
forms a well posed eigenvalue problem.  
The spectrum  should have, in principle, a lot of similarities with that   
of QNM of black holes \cite{Andersson92}, but since the boundary 
conditions at the center of the star are different than 
those at the horizon 
for a black-hole, the two spectra should not be identical. 
Nevertheless, the characteristic times of the perturbation described by 
the wave equation in the ICA will be proportional to the wave speed which 
in this case is the speed of gravitational waves and thus the perturbations 
should be of higher frequency and they should be damped out very fast 
(we remind you that for the fluid pulsation the characteristic time was 
related to the speed of the acoustic waves). 
In this way we reveal a new family of modes which is related 
directly to the spacetime and which can never be seen through 
the Newtonian  approach  to the problem.  
 
It took us nearly 20 years (from the original work by Thorne)
to recognize this natural GR extension of the stellar pulsation spectrum. 
This approach has initiated ten years ago when Kokkotas and Schutz  
\cite{KS86} have studied the QNMs of a simple ``toy" model 
consisting of two   
strings, one finite and one semi-infinite coupled together with 
a spring; this  system mimics very well the star-spacetime system. 
They have found that there   
exist two families of modes, one with slow damping (like the fluid modes)   
which were mainly modes of the finite string (star) 
and another family with 
very fast damping which was mainly modes of the semi-infinite string  
(spacetime). 
The damping time of the QNMs of this second family was shorter as the 
coupling (the spacetime curvature) of the two strings were looser.   
Such instructive ``toy" models can now easily be constructed and all of 
them suggest the existence of this second family of modes  
\cite{kokkotas85, BS93, Nils96}.  

%%%%%%%%%%%%%%%%%%%%%%%%%%  FIGURE for the w-modes %%%%%%%%%%%%%%%%%%%%%% 
\begin{figure}[htb]   
\epsfysize=9cm  
\epsfxsize=11cm 
\epsffile{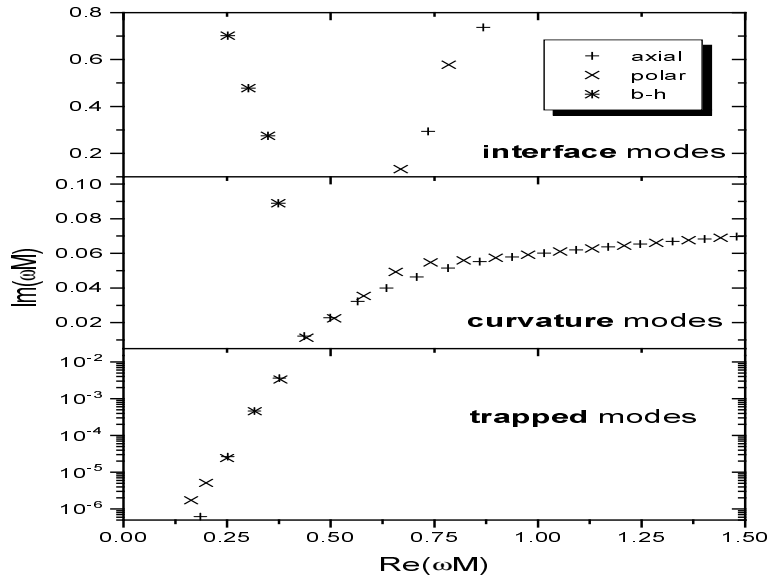}  
\caption{A graph which shows all the $w$-modes: 
{\em curvature}, {\em trapped} and {\em interface} both for axial and polar 
perturbations for a very compact uniform density star with 
$M/R=0.44$. The black-hole spectrum is also drawn for comparison. 
As the star becomes less compact the number of {\em trapped} 
modes decreases 
and for a typical neutron star ($M/R=0.2$) they disappear. 
The $Im(\omega)=1/damping$ of the {\em curvature} modes increases with 
decreasing compactness and for a
typical neutron star the first curvature mode nearly coincides with the 
fundamental black-hole mode. 
The behaviour of the {\em interface} modes slightly changes with varying 
compactness (cf. Figure 3). 
The similarity of the axial and polar spectra is apparent. 
} 
\label{modes}   
\end{figure}   
%%%%%%%%%%%%%%%%%%%%%%%%%%%%%%%%%%%%%%%%%%%%%%%%%%%%%%%%%%%%%%%%%%%%%%%%% 
   
A few years later this new family of spacetime modes has been found   
by the same authors \cite{KS92}. 
They named these new modes ``gravitational $w$ave" modes or $w$-modes.   
\footnote{ 
It should be mentioned that Kojima \cite{Koj88} had earlier found a few of   
the $w$-modes but he had not revealed the whole spectrum and 
its properties. 
} 
The characteristic properties of these modes, as it was expected by the   
previous analysis and the ``toy" model, were:   
(a) {\em high frequencies} (8-12 kHz),  
(b) {\em fast damping times} (0.02-0.1 msecs) which decrease with the order  
    of the mode and/or with the compactness of the star 
    (as the ``toy" model predicted), and 
(c) practically {\em no significant fluid motions} inside the star. 
  
The $w$-modes exist for both polar and axial perturbations since 
they do not  
depend on the perturbations of the fluid, actually in the black-hole  
perturbations the spectra in the two cases coincide. Nevertheless, 
for stars the boundary conditions near the center are different 
than those for black-holes 
(as well as the form of the potential inside the star) 
and naturally we should not  
expect coincidence, but instead a similarity of the axial 
and polar $w$-mode spectra as it has been shown in \cite{KAK95, AKK96, AKS96}, 
see also Figure \ref{modes}. 
So in what follows we shall not discriminate the axial and polar  
perturbations.

A way to understand the nature of these modes is the ``trapping" of  
the gravitational waves by the spacetime curvature inside the star.  
It is easy to see how this may happen if one plots the gravitational-wave  
speed as measured by an observer at infinity: 
$g_{tt}$ as a function of $r$.   
This has a minimum at the center of the star, and the interior   
$w$-modes would be trapped in this ``bowl'' of curvature.   
Moreover, such modes would naturally be concentrated 
at the center of the star,  
which agrees well with the  eigenfunctions constructed in \cite{LNS93}   
(see also an extensive discussion in \cite{AKK96}).  
Hence, it makes sense to refer to these modes 
as ``{\em curvature modes}''.  
  
A naive but useful argument leads to standing  
wave solutions inside the star, essentially of the 
form $\sin(\omega r)$.   
If this were the true form of the solutions, and the modes 
only leak out slowly  through the surface, we would get  
\begin{equation}  
\omega_n R =  n \pi \quad , \quad n = 1,2,...  
\end{equation}     
where $R$ is the stellar radius.
This argument is far too simplistic, but it is rewarding to find that  
two of its predictions agree well with the results for the curvature modes:  
There would exist  
an infinite sequence of such modes, and their pulsation frequencies  
${\rm Re} (\omega_n R)$ would be separated by $\pi$   
(cf. Figure 5 in \cite{AKK96}).   
This dependence on the size of the star was also predicted by the simple 
``toy" model in \cite{KS86}. 
 
When the star is made increasingly compact, 
a few of the {\em curvature} modes 
clearly change character and should be considered as trapped in the  
potential well that arises inside the black-hole potential barrier ($R<3M$).    
In this regime even the fluid $f$-mode should also 
be considered as a trapped  
mode and shares the properties of the $w$-modes 
(cf. Figure 1 in \cite{AKK96}). 
It will be instructive to name all the modes which are 
trapped because of the  
potential well {\em trapped} spacetime modes. 
Their number is finite and depends on the depth of the potential well.  
The existence of trapped modes  
has been first discussed by Detweiler \cite{Det75} but the interest  
on them has been revived with the work of Chandrasekhar and Ferrari  
\cite{CF91b} for the axial stellar perturbations.   
The existence of {\em trapped} modes for both 
polar and axial perturbations  
has been only recently shown \cite{KAK95, AKK96, AKS96} but they are  
unlikely to be of any great astrophysical relevance, such compact stars 
will probably never form.  
 
%%%%%%%%%%%%%%%%%%%%% FIGURE for the CURVATURE modes %%%%%%%%%%%%%%%%%%%%% 
\begin{figure}[htb] 
\epsfysize=7cm 
\epsfxsize=11cm
\centerline{\epsffile {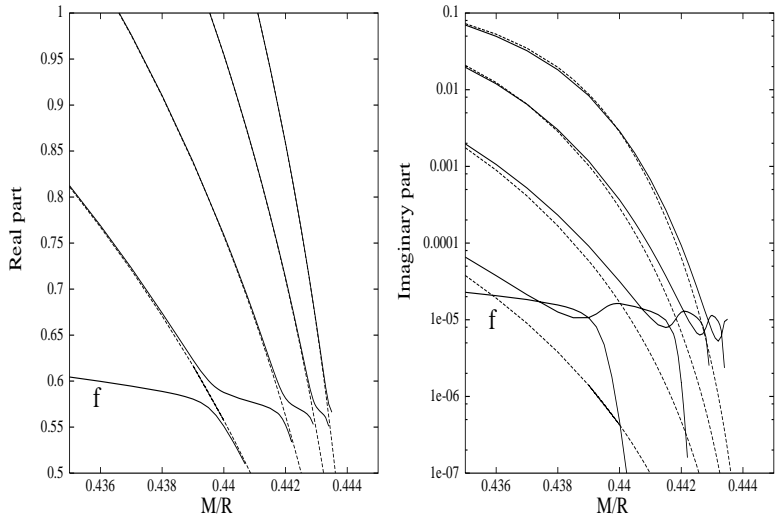}} 
\caption{ 
(left) ${\rm Re}\ \omega (R^3/M)^{1/2}$ and (right) ${\rm 
Im}\ \omega (R^3/M)^{1/2}$ as functions of the compactness of the star 
for the $f$-mode (solid and denoted by f), and the first few 
$w$-modes for polar (solid) and axial (dashed) perturbations. 
Notice the beautiful example of avoided crossings in the 
pulsation frequencies. 
} 
\label{curvmo} 
\end{figure}
\begin{figure}[htb] 
\epsfysize=7cm  
\epsfxsize=10cm
\centerline{\epsffile {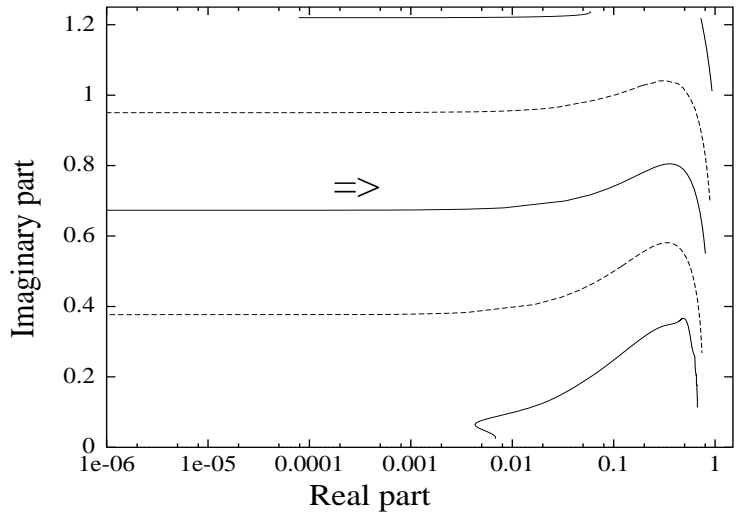}}  
\caption{ 
${\rm Im}(\omega M)$ vs ${\rm Re}(\omega M)$ for the 
polar (solid) and  axial (dashed) interface modes. Only the first of these  
modes exist for all values of  stellar compactness. Further modes arise  
for sufficiently compact stars. 
The numerical calculation becomes  
difficult for large ${\rm Im}(\omega M)$ and consequently we have only 
partial data for the third of the polar modes modes. 
Arrows indicate the direction of  increasing compactness. 
} 
\label{interfmo} 
\end{figure}

 
%%%%%%%%%%%%%%%%%%%%%%%%%%%%%%%%%%%%%%%%%%%%%%%%%%%%%%%%%%%%%%%%%%%%%%%%% 
The other family of $w$-modes discovered by Leins et al. \cite{LNS93}  
was not predicted before by any ``toy" model, and it is characterized  
by the extremely fast damping and the quite low frequencies;   
they named them $w_{II}$-modes.   
The existence of this  family of $w$-modes may be more directly  
due to the discontinuity at the surface of the star. 
Then the mode-eigenfunctions need not be localized in the star. 
Rather, these modes would be similar to modes for acoustic waves scattered 
of a hard sphere. 
One would typically expect such modes to be short-lived 
compared to modes trapped inside the star. 
Only a finite number of modes  
exist for each multipole $\ell$, and there may be purely imaginary ones.  
The latter feature suggests that one should perhaps not be  
surprised to find stellar modes ``emerge'' from points on the  
imaginary-frequency axis as in Figure \ref{interfmo} here. 
This evidence seems to be compelling and makes the association of the  
second family of $w$-modes and the interface 
at the stellar surface likely.  
Hence, it will be instructive to call these modes {\em interface modes}.  
In principle, the conjecture that  
only a finite number of these modes exist should be testable. But at   
present numerical difficulties restrict calculations   
to ${\rm Im}(\omega M) < 1.25$ or so. 
Much better numerical schemes, or other formulations of the problem,   
are required to test this prediction.  

%%%%%%%%%%%%%%%%%% FIGURE for the INTERFACE modes %%%%%%%%%%%%%%%%%%%%%%%% 
%%%%%%%%%%%%%%%%%%%%%%%%%%%%%%%%%%%%%%%%%%%%%%%%%%%%%%%%%%%%%%%%%%%%%%%%% 
 
Finally, we shall discuss some of the new results induced by 
the slow rotation of neutron stars. 
As we have mentioned earlier the eigenfrequencies are degenerate 
with respect to $m$ at fixed $\ell$. 
In the presence of rotation, however, the degeneracy with respect to $m$ 
is removed and different modes are mixed with each other. 
The frequency and damping times of the co-rotating $f$-mode 
increase with the stellar angular velocity while those of the 
counter-rotating  
mode decrease \cite{Koj93}. 
Moreover, the axial mode with $\ell$,$m$ is  
coupled with the polar mode with $\ell \pm 1$, $m$ and vice versa  
\cite{CF91c, Koj92, Koj93a}. This results from the 
dragging of the inertial  
frames (Lense-Thirring effect) by the rotation of the star.

\section{EXCITATION AND PARAMETER ESTIMATION}  
  
In the previous sections we have seen all the new developments in our  
understanding of the role of the spacetime for pulsating stars.  
Nevertheless, many important questions remain, and the most 
important of all of these 
is whether the $w$-modes can contribute to observable  
gravitational waves, and thus play a role in astrophysics.  
To test this, Andersson and Kokkotas  \cite{AK96} have studied scattering of  
axial gravitational wave-packets by a compact star.  
The problem is similar to the black hole one that was studied by 
Vishveshwara in early `70s  \cite{vishu}.  
The axial problem corresponds to a single wave equation with an  
effective potential.  
It is remarkable that what must be considered a basic exercise 
in numerical analysis can provide us with new and important information.  
  
\begin{figure}[htb]  
\epsfysize=10cm  
\epsfxsize=13cm 
\centerline{\epsffile {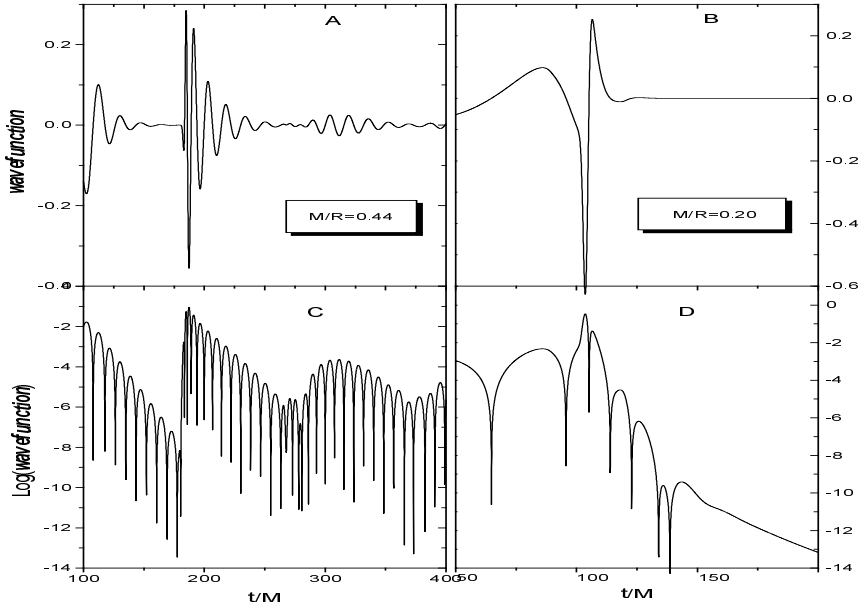}}  
\caption{ The response of two uniform density stars ($M/R=0.44$ left and   
$M/R=0.20$ right) to a Gaussian pulse of gravitational waves. 
The top shows   
the actual form of the axial perturbation as seen by a distant 
observer, while   
the lower panel shows the same function on a logarithmic scale.   
The QNM ringing that is apparent after $t\approx 100 M$ (right)   
corresponds to the first axial {\em curvature} $w$-mode, while there are   
no {\em trapped} $w$-modes. The power-law tail dominates after  
$t\approx 150M$. In the left picture we can see a series of QNMs which   
correspond both to {\em curvature} and {\em trapped} $w$-modes.  
}   
\label{waveform}  
\end{figure}

The result of a typical simulation is shown in Figure \ref{waveform}.  
The exponential ring-down at late times  (from $t\approx 100M$)   
corresponds to the  first axial $w$-mode (Figure \ref{waveform}B-D).  
A theoretical waveform based on the slowest damped  
axial $w$-mode for this star fits the numerical one perfectly.  
After the QNM ring-down (at very late times) the waveform is dominated  
by a power-law tail.  
This is exactly what one would expect \cite{waimo}.  
It is important to be mentioned that the last part of the 
signal would be the  
same if the star was replaced by a black hole of the same mass. 

%%%%%%%%%%%%%%%%%%%%%%  FIGURE for EXCITATION %%%%%%%%%%%%%%%%%%%%%%%%%%% 
%%%%%%%%%%%%%%%%%%%%%%%%%%%%%%%%%%%%%%%%%%%%%%%%%%%%%%%%%%%%%%%%%%%%%%%%% 
  
Similar simulations for ultracompact stars ($R<3M$) have been also 
performed (Figure \ref{waveform}A-C), 
for these there will be a few {\em trapped} modes. 
In this case we generally find that the first mode  
``above the peak'' [ with $(\omega M)^2> V_{\rm max}$, this is not a  
{\em trapped} mode] dominates the radiation initially.  
Most of the energy goes into this mode, but since it is more rapidly damped  
than the trapped ones, the latter dominate the radiation at later times.  
But as we mentioned earlier the ultracompact stars 
are interesting only from a  
theoretical point of view: We can learn little astrophysics from them. 
\footnote{ 
Recently, Borelli and Ferrari \cite{BF96} came to similar results about 
the excitation of the {\em trapped} axial modes by a mass falling onto 
a compact star. 
Their method is similar to the one employed by Kojima \cite{Kojima87} 
for the study of stellar resonances due to orbiting particles. 
} 
An extension of the above work is to study the excitation of both fluid and 
spacetime modes, {\em i.e.} polar perturbations, there one can exactly 
determine the amount of energy radiated due to fluid ($f$, $p$-modes) and 
spacetime modes  \cite{AAKS96}.

The simulations suggest that the $w$-modes will be relevant 
in many dynamical processes involving neutron stars.  
The modes ought to be excited  when a stellar core collapses to form  
a neutron star.  
Much of the initial deformation of spacetime could then be radiated away in  
terms of $w$-modes. 
The modes should also be excited when two neutron stars collide or at the  
final stages of binary neutron star coalescence.  
   
\begin{figure}[htb]  
\epsfysize=7cm  
\centerline{\epsffile {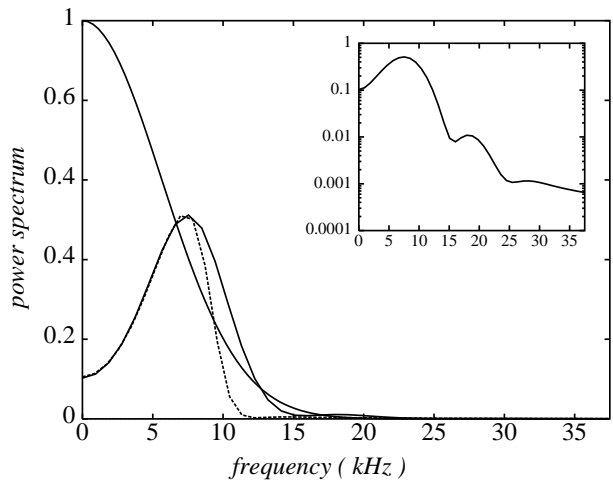}}  
\caption{ 
The power spectrum for the simulation in Figure \ref{waveform}B-D.  
The axial power  
spectrum (solid line) is compared to the corresponding 
one for a black hole   
(dashed line). We also show the power in the initial Gaussian pulse   
(another solid line) that has been used to normalize the other spectra.   
Note that the star mode is excited to roughly the same 
level as the black-hole   
one. A significant part of the initial energy clearly goes into  
quasinormal-mode ringing.  
The insertion shows the axial power spectrum in logarithmic scale.  
From this it is clear that the first three $w$-modes are excited 
(the corresponding frequencies are $\omega M \approx 0.35$, 0.82 and 1.25  
which corresponds to $\omega \approx 8.36$, 19.6 and 29.8 kHz)}  
\label{spectral}  
\end{figure}

It seems clear that the $w$-modes can play a r\^ole in many astrophysical  
scenarios, but will we be able to observe them with future 
gravitational-wave  
detectors? This is a key question, the answer of which demands much more  
detailed calculations than the ones presented here and first 
of all 3D fully  
GR codes for gravitational collapse or fully GR evolution 
codes for the binary  
neutron star coalescence.  
But let us nevertheless suggest a handwaving argument:  
A spectral analysis (see Figure \ref{spectral})  
shows that the $w$-modes will be excited to roughly the same   
level as the modes of a black hole would be in a similar situation.    
The black-hole modes are expected to be detectable \cite{finn92},   
so the situation looks promising also in the case of stars.   

%%%%%%%%%%%%%%%%%%%%%%% FIGURE for the POWER SPECTRUM %%%%%%%%%%%%%%%%%%%% 
%%%%%%%%%%%%%%%%%%%%%%%%%%%%%%%%%%%%%%%%%%%%%%%%%%%%%%%%%%%%%%%%%%%%%%%%%% 
  
In fact, the frequencies of the $f$-modes (around 1-2 kHz) make them well 
suited for detection by the operating resonant bar detectors and the 
recently proposed spherical solid-mass detectors, 
{\em e.g.} TIGA \cite{tiga1,tiga2}. 
The detection of the $w$-modes (and even of $p$-modes) 
which have frequencies 
of the order of 8-12 kHz will be possible by the proposed arrays of small 
resonant detectors \cite{FPB95}.  
If a substantial fraction of the binding energy of a neutron star   
were released in $w$-modes then these detectors could well see  
such events out to the Virgo cluster.  
  
\begin{figure}[htb]  
\epsfysize=10cm  
\epsfxsize=14cm 
\centerline{\epsffile {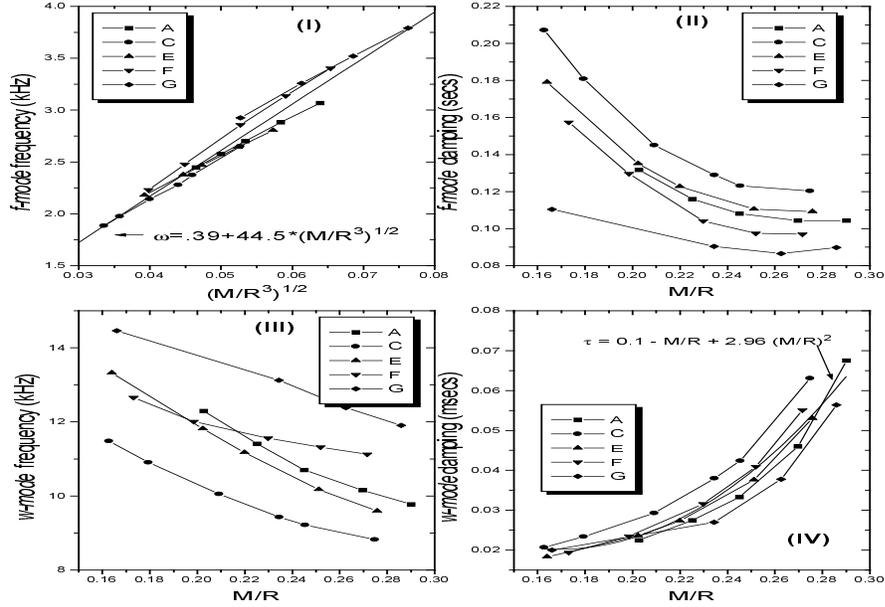}}  
\caption{ 
(I) The pulsation frequency of the $f$-mode as a function of the mean  
    density of the star,  
(II) the damping of the same $f$-mode as function of the 
     stellar compactness, 
(III) the frequency of the slowest damped $w$-mode 
     as function of the stellar  
    compactness and  
(IV) the damping rate of the same $w$-mode  as a function of the stellar  
     compactness.  
The symbols stand for the corresponding EOS {\em i.e.} 
  A for Pandharipandi (1971), 
  C for Bethe and Johnson (1974), 
  E for Moszkowski (1974), 
  F for Arponen (1972) and 
  G for Canuto and Chitre (1974). 
  }  
\label{fit}  
\end{figure}

Suppose that we detect a gravitational-wave  
signal from a compact star, what can we hope to learn from it?   
In what follows we shall suggest a possible scheme that looks very  
promising.  
The idea is the following: Assume that we  detect a signal and manage  
to extract both the fundamental polar $w$-mode and the fluid $f$-mode  
from it.  
The discussion in the previous two sections suggests that:  
\begin{itemize}  
\item the oscillation frequency of the $f$-mode scales with the 
      mean density  
      of the star $\sqrt{M/R^3}$ (cf. Figure \ref{fit}-I), 
\item the damping rate of the $w$-mode scales with the compactness ratio  
      of the star $M/R$ \cite{KS92} (cf. Figure \ref{fit}-IV). 
\end{itemize}  
  
%%%%%%%%%%%%%%%%%%%%%%%%  FIGURE for mode FITTING  %%%%%%%%%%%%%%%%%%%%% 
%%%%%%%%%%%%%%%%%%%%%%%%%%%%%%%%%%%%%%%%%%%%%%%%%%%%%%%%%%%%%%%%%%%%%% 
  
These properties are illustrated in Figure \ref{fit}.  
In principle, one should therefore be able to infer \underline{both}   
the mass and the radius of the star from the observed data.  
Additionally, the information from the damping time of the $f$-mode  
(Figure \ref{fit}-II) and the frequency of a 
$w$-mode (Figure \ref{fit}-III) 
will pose strict constraints on the possible neutron star EOS.  
We are not aware of any other scheme that would enable us to  
extract such a precise information, and put as strong constraints on the  
nuclear equation of state. 
Moreover, if the star is rotating   
the spectrum changes and the degree of deviation   
from the non-rotational case could help us to estimate the rotation 
rate of the star just as for black-holes \cite{finn92}.  
  
The idea seems simple enough, but will it be useful in practice?  
We have constructed a number of different neutron star models 
with realistic  
EOS from those listed in \cite{LD83} and we 
have determined the $f$-mode and the slowest damped  
polar $w$-mode for each of these models.  
The relevant data are graphed in Figure \ref{fit}.  
The theoretical spectra are well fitted by the following two relations:  
The oscillation frequency varies with the average density of the star as  
\begin{equation}  
\omega (kHz) = 0.39 + 44.45 \left({M\over R^3}\right)^{1/2}   
\label{detec1}   
\end{equation}   
while the damping rate of the $w$-mode behaves as 
\begin{equation}   
\tau (msec) = 0.1 - {M\over R} + 2.96 \left( {M\over R}\right)^2   
\label{detec2}   
\end{equation}   
($M$, $R$ in kms). 
If the two relations (\ref{detec1}) and  
(\ref{detec2}) provide reasonably accurate estimates for both $M$ and $R$  
for all stars in our dataset, then this idea passes a first test.  
The procedure passes this simple test with flying colours:  
The mass and the radius of each star can be determined to within 5\%  
by inverting (\ref{detec1}) and (\ref{detec2}).  
The results of such a test are shown in Table \ref{Table1}, where we have  
tested whether the above relations can estimate the masses and radii of 
various polytropes.  
  
The evidence provided here is, of course, only an  
indication that this idea can be useful in a real detection situation. 
To investigate this in more detail one must incorporate the estimated  
effect of statistical errors and measurement ones. 
It will, for example,  
be much more difficult to infer the $w$-mode damping rate from a 
data set than  
to find the $f$-mode pulsation frequency (the observability of a periodic  
signal buried in noise scales roughly as the square-root of the number  
of observed cycles), and 
also to obtain fits similar to (\ref{detec1}) and (\ref{detec2})  
for all known realistic EOS. 
If such relations prove as robust as the ones we have obtained 
here then the suggested scheme looks truly promising. 
In principle it will be possible  
even to identify exotic type of stars like the boson stars which 
also show a $w$-mode spectrum \cite{futam94}. 
%%%%%%%%%%%%%%%%%%%  TABLE  for the comparison %%%%%%%%%%%%%%%%%%%%%%%%%%% 
\begin{table}   
\caption{   
Test of the estimation of parameters hypothesis on polytropic models with  
N=0.8, 1 and 1.2, the values in 
parentheses are the estimated values from the relations (\ref{detec1}) and   
(\ref{detec2}).} 

\begin{tabular}[t]{cccccc}   
\hline   
\hline   
N &$R$ (Km) & $M/M_\odot$ & $M/R$ & $\omega_f$ (kHz) & $\tau_w$ (msecs) \\  
\hline   
0.8 & 10.21 (10.54) & 0.92 (0.97) &0.133 (0.135)& 1.933 & 0.0152 \\   
0.8 &  9.49 (9.61)  & 1.36 (1.35) &0.211 (0.207)& 2.532 & 0.0238 \\   
 1  &  8.86 (8.35)  & 1.27 (1.16) &0.206 (0.211)& 2.870 & 0.0236 \\   
 1  &  7.42 (7.28)  & 1.35 (1.31) &0.266 (0.269)& 3.668 & 0.0493 \\   
1.2 & 10.48 (9.89)  & 1.44 (1.48) &0.203 (0.221)& 2.540 & 0.0276 \\   
1.2 &  8.97 (9.45)  & 1.46 (1.45) &0.240 (0.254)& 3.128 & 0.0422 \\   
\hline   
\hline   
\label{Table1}   
\end{tabular}   
\end{table}

\section{CONCLUSIONS}   
  
In this review we have shown that the study of the stellar pulsations 
as sources of gravitational waves should be performed in GR 
since  in the Newtonian gravity 
apart from deviations in the strong field regime,  
a lot of useful physical information is overlooked. 
 
The existence of new families of modes (not seen in Newtonian  
theory) which are not just another branch of the possible mathematical  
solutions of the eigenvalue problem but instead are modes 
which can be excited  
and detected enrich the information that we can get from the signals of  
pulsating stars.  
 
The assertions presented here must be tested 
by more detailed, fully general relativistic simulations.  
This is, in fact, an interesting challenge for numerical relativity.  
A relativistic description of gravitational collapse to form a 
neutron star, or the merger of two stars,  should tell us whether the 
$w$-modes are of observational relevance or not.  
Given the present importance for TIGA, fully relativistic simulations are  
urgently required. 
 
Moreover, there is still a lot of work to be done on the understanding of  
stellar pulsations. 
For example, we have not discussed   the  rotating case,  
which in GR  is tractable only in the limit of slow rotation.  
Thus a natural question  is:
do the $w$-modes induce new instabilities or  
enchance the classical gravitational radiation instability for fast 
rotating stars? 
The answer to this question is of great importance for the 
gravitational wave  
astronomy and, due to the lack of full GR rotating star solutions, 
one should try to attack the problem  via pN approach \cite{GS90}.

\ack{  
I acknowledge the helpful suggestions and stimulating discussions with 
Gerhard Sch\"afer which have improved the form of this article.  
The collaboration with Nils Andersson, Yasafumi Kojima and Bernard
Schutz has greatly enhanced my understanding of this field.
This work was supported by an exchange programme from the British Council 
and the Greek GSRT and also by a visiting fellowship from the German DAAD. 
A visit at the Max-Planck Research-Unit for Gravitational Theory at the
University of Jena during which this work has been completed is gratefully 
acknowledged.
}


\begin{thebibliography}{100} 
\bibitem{BM94} 
     Bonazzola S., Marck J.A., 
     \review  Annu. Rev. Nucl. Part. Sci., 45, 1994, 655. 
\bibitem{TC67}   
     Thorne K.S., Campolattaro A.,   
     \review  Astroph. J., 149, 1967, 591.   
\bibitem{Thorne68}   
     Thorne K.S.   
     \review Phys. Rev. Let., 21, 1968, 320.   
\bibitem{PT69}   
     Price R.H., Thorne K.S.,   
     \review Ap. J., 155, 1969, 163.   
\bibitem{Thorne69a}   
     Thorne K.S.,   
     \review Ap. J., 158, 1969, 1.   
\bibitem{Thorne69b}   
     Thorne K.S.,   
     \review Ap. J., 158, 1969, 997.   
\bibitem{KS86}   
      Kokkotas K.D., Schutz B.F.,   
      \review G.R.G., 18, 1986, 913.   
\bibitem{KS92}   
      Kokkotas K.D., Schutz B.F.,   
      \review M.N.R.A.S, 255, 1992, 119.   
\bibitem{CF91a}   
      Chandrasekhar S., Ferrari V.,   
      \review Proc. R. Soc. London A, 432, 1991, 247.   
\bibitem{Koj88}   
      Kojima Y.,   
      \review Progr. Theor. Phys., 79, 1988, 665.   
\bibitem{AKS95}   
      Andersson N., Kokkotas K.D., Schutz B.F.   
      \review  M.N.R.A.S., 274, 1995, 1039.   
\bibitem{KAK95}   
      Kojima Y., Andersson N., Kokkotas K.D.,   
      \review Proc. R. Soc. Lond. A, 471, 1995, 341.   
\bibitem{AKK96}   
       Andersson N., Kojima Y., Kokkotas K.D.,   
       {\em Ap. J.} in press (1996), gr-qc/9512048.    
\bibitem{AKS96}   
      Andersson N., Kokkotas K.D., Schutz B.F.   
      {\em Spacetime modes of relativistic stars}, 
      {\em M.N.R.A.S.} in press (1996), gr-qc/9601015   
\bibitem{LNS93}   
       Leins M., Nollert H-P., Soffel M.H.,   
      \review Phys. Rev. D, 48, 1993, 3467.   
\bibitem{FG94}  
      Ferrari V., Germano M.,  
      \review Proc. R. Soc. Lond. A, 444, 1994, 389.  
\bibitem{CF95}   
      Chandrasekhar S., Ferrari V.,   
      \review Proc. R. Soc. Lond. A, 450, 1995,  463.   
\bibitem{CF91c}   
      Chandrasekhar S., Ferrari V.,   
      \review Proc. R. Soc. Lond. A, 433, 1991, 423.   
\bibitem{Koj92}   
      Kojima Y.,   
      \review Phys. Rev. D, 46, 1992, 4289.   
\bibitem{Koj93}  
      Kojima Y.,  
      \review Ap. J., 414, 1993, 247.  
\bibitem{Koj93a}  
      Kojima Y.,  
      \review  Prog. Theor. Phys.,  90, 1993, 977.   
\bibitem{CF91b}   
      Chandrasekhar S., Ferrari V.,   
      \review  Proc. R. Soc. Lond. A, 434, 1991,  449.   
\bibitem{KDK94}   
      Kokkotas K.D.,   
      {\em  M.N.R.A.S.} {\bf 268} (1994) 1015; 
      {\em  M.N.R.A.S.} {\bf 277} (1995) 1599.   
\bibitem{AK96}   
      Andersson N., Kokkotas K.D., 
      {\em Gravitational waves and pulsating stars: 
      What can we learn from future observations?} 
      preprint (1995). 
\bibitem{Schutz87} 
      Schutz B.F., 
      in {\em Gravitation in Astrophysics}, 
      ed. B. Carter and J.B. Hartle (New York: Plenum Press) (1987). 
\bibitem{DL85} 
      Detweiler S.L.,  Lindblom L., 
      \review Ap. J.,  292, 1985, L12. 
\bibitem{IP91} 
      Ipser J.R.,  Price R.H., 
      \review Phys. Rev. D,  43, 1991, 1768. 
\bibitem{Moncr74} 
       Moncrief V. 
       \review   Ann. Phys.,  88, 1974,  343.  
\bibitem{Chandra83} 
      Chandrasekhar S., \book The mathematical theory of black holes,   
      Oxford University Press, New York, 1983, 144-150.   
\bibitem{Cowling41}   
      Cowling T.G.,   
      \review M.N.R.A.S., 101, 1941, 367.   
\bibitem{Kelvin}  
       Thomson W.,  
       \review Phil. Trans. Roy. Soc. London, 153, 1863, 603. 
\bibitem{LD83}   
       Lindblom L., Detweiler S.L.,   
       \review  Ap. J. Suppl.,  53, 1983, 73.   
\bibitem{ST83}   
      Schumaker B.L., Thorne K.S.,   
      \review M.N.R.A.S., 203, 1983, 457.   
\bibitem{finn90}  
        Finn L.S.,  
       \review  M.N.R.A.S., 245, 1990, 82.  
\bibitem{McD88}   
      McDermott P.N., Van Horn H.M., Hansen C.J.,   
      \review  Ap. J., 325, 1988, 725. 
\bibitem{OH73} 
      Osaki Y., Hansen C.J.,
      \review   Ap. J., 185, 1973, 277.  
\bibitem{BS82}   
      Balbinski E., Schutz B.F.,   
      \review  M.N.R.A.S.,   200, 1982, 43P.   
\bibitem{BDLS85}   
     Balbinski E.,  Detweiler S., Lindblom L., Schutz B.F.,   
     \review M.N.R.A.S., 213, 1985, 553.   
\bibitem{cutler91}  
      Cutler C.,  
      \review Ap. J., 374, 1991, 248.  
\bibitem{CL92}  
      Cutler C., Lindblom L.,  
      \review Ap. J., 385, 1992, 630.              
\bibitem{KS96}   
      Kokkotas K.D., Sch\"afer G.,    
      {\em Post-Newtonian Stellar Oscillations}, 
      preprint (1996)   
\bibitem{CLS90}  
      Cutler C., Lindblom L., Splinter R.J.,  
      \review Ap. J., 363, 1990, 603.        
\bibitem{Robe68}   
       Robe H.,   
       \review  Annales d' Astrophysique, 31, 1968, 475.   
\bibitem{McDer83}  
       McDermott P.N., Van Horn H.M., Scholl J.F.,  
       \review  Ap. J.,  268, 1983,  837.  
\bibitem{finn88}  
        Finn L.S.,  
       \review  M.N.R.A.S., 232, 1988, 259.  
\bibitem{Unno89}   
       Unno W., Osaki Y., Ando H.,  Saio H., Shibahashi H.,  
       \book  Nonradial Oscillations of Stars,  
       University of Tokyo Press, Japan, 1989, 113.  
\bibitem{Andersson92}   
      Andersson N.,   
      \review Proc. R. Soc. London A,  439, 1992, 47.   
\bibitem{kokkotas85} Kokkotas K.D.,  
      {\em MSc Thesis, University of Wales}, (1985)  
\bibitem{BS93}  
      Baumgarte T., Schmidt B.G.,  
      \review Class. Quantum Grav., 10, 1993, 2067.  
\bibitem{Nils96}  
       Andersson N.,  
%       {\em Two simple models for gravitational-wave modes of 
%       compact stars}  
       preprint (1996)         
\bibitem{Det75}  
      Detweiler S.L.,  
      \review Ap. J., 197, 1975, 203.  
\bibitem{vishu}  
      Vishveshwara C.V.,  
      \review Nature, 227, 1970, 936.  
\bibitem{waimo}  
      Ching E.S.C., Leung P.T., Suen W.M., Young K.,  
      \review  Phys. Rev. D, 52, 1995, 2118.  
\bibitem{BF96} 
      Borrelli A., Ferrari V., 
      {\em preprint} (1995). 
\bibitem{Kojima87}   
      Kojima Y.,   
      \review Progr. Theor. Phys., 77, 1987, 297.   
\bibitem{AAKS96}  
      Allen G., Andersson N., Kokkotas K.D., Schutz B.F.,  
      {\em work in progress}.  
\bibitem{finn92}  
      Finn L.S.,  
      \review Phys. Rev. D, 46, 1992, 5236.  
\bibitem{Pand71}  
        Pandharipande V., 
        \review Nucl. Phys. A, 178, 1971, 123. 
\bibitem{BJ74} 
        Bethe H.A., Johnson M.,  
        \review Nucl. Phys. A, 230, 1974, 1. 
\bibitem{Moszk74} 
        Moszkowski S.,  
        \review Phys. Rev. D, 9, 1974, 1613.  
\bibitem{Arpon72} 
         Arponen J., 
        \review Nucl. Phys. A, 191, 1972, 257.   
\bibitem{CC74} 
        Canuto V., Chitre S.M., 
        \review Phys. Rev. D, 9, 1974, 1587.          
\bibitem{tiga1}  
        Johnson W.W., Merkowitz S.M.,  
        \review Phys. Rev. Letters, 70, 1993, 2367.  
\bibitem{tiga2}  
        Merkowitz S.M., Johnson W.W.,  
        \review Phys. Rev. D, 51, 1995, 2546.  
\bibitem{FPB95} Frasca S., Papa M.A., Bassan M.,
       {\em preprint} Rom2F/94/20.         
\bibitem{futam94}  
       Yoshida S., Eriguchi Y., Futamase T.,  
       \review Phys.Rev. D, 50, 1994, 6235.  
\bibitem{GS90}
       Sch\"afer G.,
       \review Astron. Nachr., 311, 1990, 213.  
   
\end{thebibliography}
\end{document}